\documentclass[twocolumn,amssymb,fleqn]{revtex4} 
\usepackage{epsfig,amssymb,amsmath,graphicx,subfigure,hyperref  }  
\usepackage{color}

\newcommand{\be}{\begin{eqnarray}}   
\newcommand{\ee}{\end{eqnarray}}

\begin{document}

\title{Shapes of pored membranes}
\author{Zhenwei Yao, Rastko Sknepnek, Creighton K. Thomas and Monica Olvera de la Cruz}
\affiliation{Department of Materials Science and Engineering,
Northwestern University, Evanston, Illinois 60208-3108, USA}
\begin{abstract}
  We study the shapes of pored membranes
within the framework of the Helfrich theory under the constraints
of fixed area and pore size. We show that the mean curvature term
leads to a budding-like structure, while the Gaussian curvature
term tends to flatten the membrane near the pore; this is
corroborated by simulation. We propose a scheme to deduce the
ratio of the Gaussian rigidity to the bending rigidity simply by
observing the shape of the pored membrane. This ratio is usually
difficult to measure experimentally. In addition, we briefly
discuss the stability of a pore by relaxing the constraint of a
fixed pore size and adding the line tension. Finally, the
flattening effect due to the Gaussian curvature as found in
studying pored membranes is extended to two-component membranes.
We find that sufficiently high contrast between the components'
Gaussian rigidities leads to budding which is distinct from that
due to the line tension.
\end{abstract}
\maketitle

\section{Introduction}

The cell membrane is a complex bilayer sheet consisting of
hundreds of lipid species embedded with numerous surface- and
trans-membrane proteins.\cite{alberts2007molecular} Its main role
is to separate the cell's interior from its surroundings and to
act as a conduit for exchanging matter and signaling between the
cell and its environment. The cell membrane is a dynamic object
whose conformational variations are associated with biological
activities such as cell fission, fusion, and
adsorption.\cite{Lip_Sack:1995} Most biological membranes exist in
a liquid state where lipid molecules are rather strongly confined
to the bilayer plane but can easily diffuse laterally within it.
Fluidity allows the membrane to dynamically rearrange its local
composition, quickly heal holes, and enable transmembrane
transport besides allowing other metabolic functions. A
substantial portion of the transport through the cell membrane
takes place via pores.\cite{Biochemstry:book,lipid_rafts:2009} The
presence of a pore changes the topology of a membrane and can
significantly influence its conformations and
functions.\cite{talin:1998} Characterizing the conformations of
closed membranes has been a subject of active research over the
past four decades with numerous
experimental\cite{Torus:1992,Xavier:1994} and
theoretical\cite{SMoMS,OuYang:1989,Lipowsky:1991} studies.

Despite a high molecular complexity, when the length scale is
large compared to the bilayer thickness and the energy scale is
small compared to the typical intermolecular interactions, the
cell membrane shape can be successfully described by a simple
model proposed by Helfrich nearly forty years
ago.\cite{Helfrich1973} In his seminal paper, Helfrich argued that
the low-energy large-scale properties of a liquid membrane can be
described in terms of a free energy that is a quadratic function
of the two principal curvatures expressed in terms of their two
invariants: the mean curvature and the Gaussian curvature. Within
the framework of the Helfrich theory, various axisymmetric and
non-axisymmetric shapes of closed membranes have been
predicted.\cite{LCMembrane:1999,Safran2003} In particular, the
longstanding physiological puzzle about the biconcave shape
typical of the red blood cell has been beautifully solved; the
shape of the red blood cell shape has been understood as the
conformation that minimizes the Helfrich free energy under a set
of prescribed volume and area
constraints.\cite{red_blood_cell:1976} Many predictions based on
the Helfrich free energy have been observed experimentally.
\cite{OuYang:1989} For example, the theoretical discovery of the
thermal repulsion between membranes, that is to prevent sticking
of cells, has been confirmed by small angle X-ray diffraction
experiment.\cite{Helfrich1973,X_ray:1978}

There are various ways to form pores on membranes {\em in vivo}
and {\em in vitro}. For example, pore-forming toxin proteins exist
in a wide range of organisms including bacteria, fungi, plant and
animal cells.\cite{Gilbert:2002} By binding at particular sites on
a membrane, toxins can create pores via oligomerizing on the
membrane surface. The pores created by toxin proteins are of
limited sizes. For example, the maximum size of the pore formed by
SecYEG on E. Coli is below $2.2-2.4\ \textrm{nm}$.\cite{pore_Sec}
Recent studies have shown that larger pores can be created on a
fluid membrane by detergents\cite{pore_exp:PNAS2001} or
submembranous protein talin.\cite{talin:1998} Note that the size
of the pore is controlled by tuning the talin concentration over
an appropriate range.\cite{pore_exp:PNAS2001} The localization of
talin mainly along the pore rim, as observed by fluorescent
labelling, is likely responsible for stabilizing the pore. A
recent experiment introduced a method to create pores of about
$15\ \textrm{nm}$ in a lipid membrane.\cite{Dubois:2001} In a
salt-free catanionic solution, charged pores are produced on
membranes due to the partial segregation of the anionic surfactant
in excess. In this case, the size of a pore can be controlled by
tuning the relative amount of anionic and cationic surfactants and
thus the charges on a pore. The size of a stable pore is
determined by the competition of the line tension energy $\gamma R$ and
the electrostatic self-energy $q^{2}/(\epsilon R)$, where $\gamma$
is the line tension, $R$ is the size of the pore, $q$ is the total
charge on the pore and $\epsilon$ is the dielectric constant of
the medium, such that $R\sim\sqrt{q^{2}/(\gamma\epsilon)}$. Both
the increase of charge and the decrease of line tension can
enlarge a pore on membrane.

In this paper, we study how a pore modifies the morphology of a
fluid membrane within the framework of the Helfrich theory. We
discuss the equilibrium solutions of the Helfrich shape equation
for fluid membranes with fixed area and pore size. In experiments
the fixed pore size constraint can be realized by introducing
stabilizing agents as discussed above. We find a budding structure
in pored membranes, dictated by the mean curvature term in the
Helfrich free energy. In studies of the conformation of closed
single-component liquid membranes, the Gaussian curvature term in
the Helfrich free energy can be omitted, as it is a constant that
does not depend on the membrane's shape. However, this is no
longer the case if pores are present. We show that the Gaussian
curvature term can significantly influence the shape of a pored
membrane by imposing a local constraint on the shape of the
membrane near the pore. The Gaussian curvature term tends to pull
the membrane outside a pore to the plane where the pore loop lies
and the membrane near the pore is flattened. This observation may
lead to a simple method to fabricate polyhedral buckled membranes
by manipulating the size and position of the pores. In addition,
we propose a scheme to find the ratio of the Gaussian rigidity and
the bending rigidity from the shape of a pored membrane. This
ratio is usually difficult to measure
experimentally.\cite{gaussian_rigidity:2012} The proposed scheme
successfully passes the test on a pored membrane generated by
Surface Evolver,\cite{brakke1992surface,evolver} and is applied on
an experimental case. Furthermore, we briefly discuss the
stability of a pore on a membrane by relaxing the constraint of
fixed pore size and adding the line tension. We find that a
budding pore may be meta-stable with very shallow energy barrier
and over a very narrow range of values of line tension. Therefore,
stabilizing agents like talin proteins in the experiment of
Ref.~\citenum{talin:1998} are essential for a stable pore on fluid
membranes. Finally, the flattening effect due to the Gaussian
curvature as found in studying pored membranes is extended to
two-component membranes. Multicomponent membranes can have a wide
variety of morphologies, as has been recently discussed for both
liquid\cite{Hu2011,demers2012curvature} and polymerized
membranes.\cite{Vernizzi2011,sknepnek2011buckling,Sknepnek2012} We
find that the flattening effect due to the Gaussian curvature can
induce budding in two-component membranes when there is
sufficiently high contrast between the components' Gaussian
rigidities. This is recognized as a domain-induced budding, but
via a mechanism that is distinct from the conventional line
tension driven budding.\cite{domain:line_tension,Kohyama2003}

\begin{figure}[tb]
\begin{center}
\includegraphics[scale=0.8]{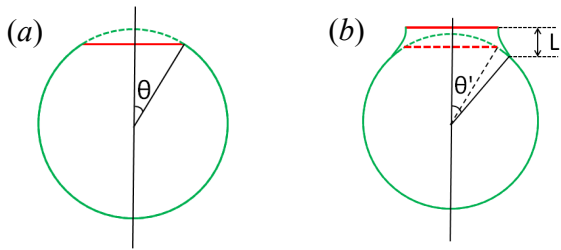}
\end{center}%
\caption{The mean curvature term in Eq.~(\ref{bending_energy})
gives rise to a budding pore (right) instead of making a membrane
spherical everywhere (left). The red line represents the opening
of the membrane.\label{why_qiao}}
\end{figure}

\section{Model}

The bending energy of a fluid membrane is modeled by the Helfrich
free energy:\cite{Helfrich1973}
\begin{eqnarray}
E=\frac{1}{2}\kappa\int\left(2H\right){}^{2}dA+\kappa_{G}\int
K_{G}dA,\label{bending_energy}
\end{eqnarray}
where $\kappa$ ($\sim 10\ k_{B}\textrm{T}$)\cite{kBT:1992} and
$\kappa_{G}$ are the bending rigidity and the Gaussian rigidity,
respectively. The mean curvature $2H=1/R_{1}+1/R_{2}$ and the
Gaussian curvature $K_{G}=1/\left(R_{1}R_{2}\right)$, where
$R_{1}$ and $R_{2}$ are the radii of principal curvatures. For
real membranes, $\kappa>0$ and $\kappa_{G}<0$.\cite{Webb:2005}
Note that in Eq.~(\ref{bending_energy}) we have assumed that the
spontaneous curvature $H_{0}=0$, as is the case if there is no
asymmetry with respect to the middle surface of the bilayer. The
negative sign of the Gaussian rigidity indicates that it favors
lower genus surfaces.\cite{Struik:1988} For example, without
considering the mean curvature term, a spherical membrane is more
stable than a toroidal membrane; the integrals of the Gaussian
curvature for sphere and torus are $4\pi$ and zero, respectively.

According to the Gauss-Bonnet theorem, the integral of the
Gaussian curvature over a manifold $M$ is related to the integral
of the geodesic curvature $k_{g}$ along the boundary of the
manifold $\partial M$ by
\begin{equation}
\int_{M}dAK_{G}=2\pi\chi\left(M\right)-\oint_{\partial M}k_{g}dl\label{GB}
\end{equation}
where $\chi(M)$ is the Euler characteristic of the manifold
$M$.\cite{Struik:1988} For a closed manifold $M$ without pores,
the geodesic curvature term vanishes and the integral of the
Gaussian curvature becomes a constant. Therefore, $\kappa_{G}$
plays no role for a topologically spherical membrane. However,
$\kappa_{G}$ becomes important if a pore is introduced into a
membrane to change its topology.\cite{Struik:1988} In fact, we
find that even without considering the Gaussian curvature term in
the Helfrich free energy, a pore on a membrane can induce an
interesting budding structure.

\section{Results and discussion}

\subsection{Single pore}

\begin{figure*}
\begin{center}
\includegraphics[scale=0.8]{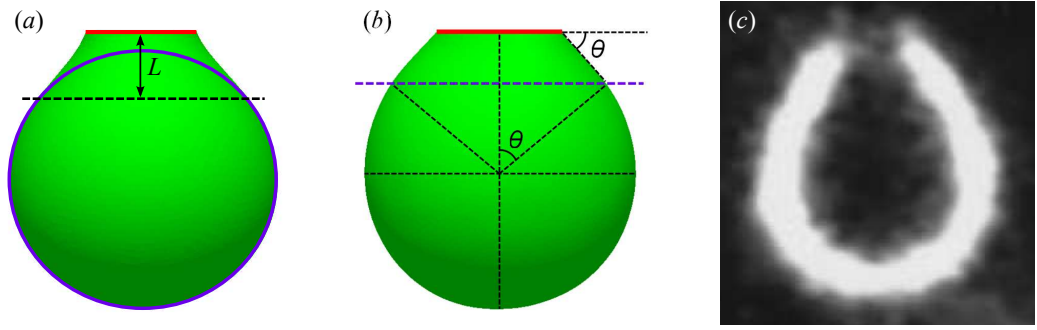}
\end{center}
\caption{The ground state shapes of a pored membrane generated by
Surface Evolver. The red line represents the opening of the
membrane. The size of the pore is fixed. In (a)
$\kappa_{G}/\kappa=0$ and in (b) $\kappa=2$ and $\kappa_{G}=-1.5$.
The comparison of (a) and (b) shows that the mean curvature term
in the Helfrich free energy leads to a budding pore, while the
Gaussian curvature term tends to flatten the membrane near the
pore. In (a), measured by the radius of the pore, the radius of
the sphere $R\approx4.95$, and the longitudinal size of the
budding pore $L\approx 1.22$, which agrees well with our
prediction (see Eq.~(\ref{LR})). (c) The shape of a pored fluid
membrane from experiment whose radius is about $1\ \mu
\textrm{m}$. The pore is created by the protein
talin.\cite{talin:1998} \label{qiao} }
\end{figure*}

\begin{figure}[h]
\begin{center} \includegraphics[width=2.5in]{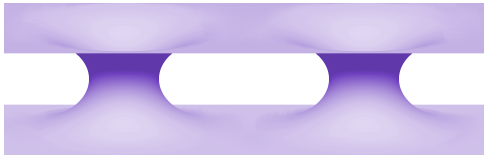}
\par \end{center} \caption{The necks between two asymptotically
flat parallel membranes repel each other as they approach; their
overlap increases the bending energy of the system.
\label{double_catenoid} }
\end{figure}

By exclusively considering the mean curvature term in the Helfrich
free energy Eq.~(\ref{bending_energy}), we analyze how the
morphology of a topologically spherical membrane is influenced by
a pore. Based on the intuition about closed membranes, one might
guess that a punctured membrane would take a spherical shape
everywhere except at the pore for minimizing the mean curvature
term in the Helfrich free energy, as in Fig.~\ref{why_qiao}(a).
Numerical experiments performed with Surface
Evolver,\cite{brakke1992surface,evolver} however, show that a budding pore appears,
as in Fig.~\ref{qiao}(a). It is thus natural to ask: Why does a
budding of the pore appear? How does such a conformation minimize
the integral of the squared mean curvature? To address these
questions, we compare the energies of the two shapes in
Fig.~\ref{why_qiao}(a, b). In order to minimize the integral of
the squared mean curvature, the neck prefers to being a minimal
surface with vanishing mean curvature. A catenoid is the only
minimal surface with rotational symmetry.\cite{Nitsche}
Consequently, the shape in Fig.~\ref{qiao}(a) is, to first
approximation, composed of a catenoid and part of a sphere
(emphasized by the purple oval in Fig.~\ref{qiao}(a)). The
integrals of the squared mean curvature of the two shapes in
Fig.~\ref{why_qiao} are calculated as:
$E_{a}=\frac{\pi}{2}\left(1+\cos\theta\right)$ and
$E_{b}=\frac{\pi}{2}\left(1+\cos\theta'\right)$, where the angles
$\theta$ and $\theta'$ are defined in Fig.~\ref{why_qiao}. Note
that the bending energy is independent of the radius of the
sphere, as the integral of the squared mean curvature is scale
invariant.\cite{SMoMS} Since $\theta'>\theta$, $E_{b}<E_{a}$, a
budding pore is preferred. Theoretical model based on the boundary
layer method shows that catenoidal necks between two
asymptotically flat parallel membranes (a wormhole like structure,
see Fig.~\ref{double_catenoid}) are interacting like a gas of free
particles with a hard core repulsion.\cite{Xavier:1994} The
repulsion between necks comes from their overlap as they approach,
which increases the bending energy of the system. It is analogous
to the capillary interaction between particles floating or
immersing on a liquid interface; their interaction originates from
the overlap of the capillary deformations near particles.\cite{Particle_Interface}
Considering that the budding pore structure is half of the wormhole like structure,
we expect these budding pores also repel each other on the membrane as they
approach.

\begin{figure}[h]
\centering{} \includegraphics[scale=1.0]{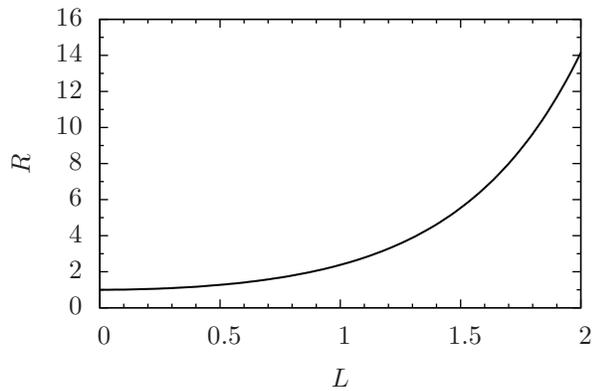}
\caption{Sphere radius, $R$, as a function of the longitudinal
size $L$ of the budding pore when $|\kappa_G| \ll \kappa$, as
given in Eq.~(\ref{eq:LR}). \label{LR} }
\end{figure}

We further calculate the longitudinal size $L$ of a budding pore,
as defined in Fig.~\ref{why_qiao}(b). We choose an x-y coordinate
system such that the x-axis is along the solid red line in
Fig.~\ref{why_qiao}(b) and the y-axis is along the symmetric axis
of the membrane. The shape of the neck is characterized by
$x\left(y\right)=r\ \cosh y$, where $r$ is the radius of the waist
of the catenoid. By assuming that the boundary of the pore falls
on the waist of the catenoid, we get the expression for the angle
$\alpha$ between the x-axis and the tangent vector at the
connecting circle of catenoid and sphere: $\cot\alpha=\sinh
(L/r)$. On the other hand, a geometric argument leads to the
relation between the radius $R$ of the sphere and the size of the
pore $L$ as $R\sin\alpha=r\ \cosh (L/r)$. From these two
expressions, we finally have
\begin{equation}
R=r\ \cosh^2 (\frac{L}{r}),\label{eq:LR}
\end{equation}
where $r$ is the radius of the pore. The dependence of the radius
$R$ of sphere on the longitudinal size $L$ of the budding pore is
plotted in Fig.~\ref{LR}. Measured in units of the radius of the
pore, $L$ increases from $1.4$ to $1.8$ as $R$ increases from $5$
to $10$. The budding of a pore is more obvious in a bigger
membrane. For the shape generated by Surface Evolver in
Fig.~\ref{qiao}(a), we measure $R=4.95$ and $L\approx 1.22$ which
is close to our prediction $L=1.4$. The deviation comes from the
assumption that the pore boundary falls on the waist of the
catenoid, which is not precisely the case in Fig.~\ref{qiao}(a).
For very large values of $R$, from Eq.~(\ref{eq:LR}), the
longitudinal size $L$ of the budding pore scales as
$L\sim\frac{1}{2}\ln R$. The logarithm function comes from the
exponential grow of the catenoidal neck from its waist.

In real fluid membranes, the Gaussian rigidity can contribute more
than $400\ \textrm{kJ}/\textrm{mol}$ in topological transformation
of a membrane like creating a pore.\cite{gaussian_rigidity:2012}
Theoretical microscopic models of monolayer fluid membranes show
that
$\kappa_G/\kappa\in\left[-1,0\right]$.\cite{templer1998gaussian}
Therefore, the Gaussian rigidity can compete with the bending
rigidity for influencing the shape of a pored membrane. In the
following, we study this problem in the light of the Gauss-Bonnet
theorem.

The Gauss-Bonnet theorem Eq.~(\ref{GB}) implies that the integral
of the Gaussian curvature can be maximized by minimizing the line
integral of the geodesic curvature, such that the bending energy
is minimized as $\kappa_{G}$ is negative. Therefore, the Gaussian
curvature term in the Helfrich free energy, which is an integral
over the whole surface, essentially imposes a local constraint on
the shape near the boundary, such that the integral of the
geodesic curvature on the boundary is minimized. The geodesic
curvature $k_{g}$ describes the deviation of a curve away from a
geodesic, a generalization of a straight line in a plane. For
example, the geodesic curvature of a big circle on a sphere is
zero, since it corresponds to a straight line on spherical
geometry. The geodesic curvature of a curve in a surface is
defined in the following way. Consider a curve $\vec{x}(s)$ being
parametrized by the arc length $s$, its curvature is
$\vec{k}=\frac{d\hat{t}}{ds}$, where $\hat{t}=\frac{d\vec{x}}{ds}$
is the unit tangent vector of the curve. For a curve on a surface
equipped with the coordinates $\left\{
\vec{e}_{u},\vec{e}_{v}\right\} $, the curvature $\vec{k}$ can be
projected along the normal and tangent plane of the surface:
\begin{equation}
\vec{k}=\frac{d\vec{t}}{ds}=\vec{k}_{n}+\vec{k}_{g},\label{eq:k}
\end{equation}
where $\vec{k}_{n}=\left(\vec{k}\right){}_{\hat{n}}$ and
$\vec{k}_{g}=\left(\vec{k}\right){}_{\textrm{TM}}$. $\vec{n}$ is
the normal vector pointing \textit{outward}; {\em i.e.}, along the
direction of $\vec{e}_{u}\times\vec{e}_{v}$. TM represents the
tangent plane. In the Gauss-Bonnet theorem, the sign of the
geodesic curvature needs to be clarified.
$k_{g}=\vec{k}_{g}\cdot\hat{u}$, where
$\hat{u}=\hat{n}\times\hat{t}$.\cite{Struik:1988} The direction of
$\hat{t}$ is chosen to be along the boundary of the pore such that
the membrane stays on the left hand side of the
boundary.\cite{Struik:1988} Under these conventions, the sign of
the geodesic curvature is unambiguously determined.

\begin{figure}
\begin{center}
\includegraphics[width=3in]{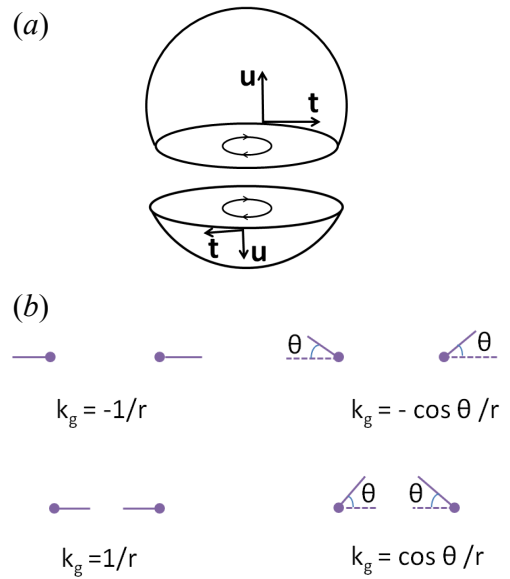}
\par\end{center}
\caption{(a) The calculation of the geodesic curvature. (b)
Possible shapes of a membrane near a circular pore which is
represented by two dots.\label{cal_kg} }
\end{figure}

Using arguments based on the Gauss-Bonnet theorem, we show that
the Gaussian curvature term in the Helfrich free energy tends to
flatten the membrane near the pore. We first calculate the
geodesic curvatures on the circular boundaries in the cut unit
sphere as in Fig.~\ref{cal_kg}(a). For the upper bigger part of
the cut sphere, the tangent vector on the boundary circle is
clockwise seen from below, so the sphere is on the left hand side
walking along the boundary circle. The other tangent vector
$\hat{u}$ points upward, as shown in Fig.~\ref{cal_kg}(a), because
the normal vector points outward. The curvature vector $\vec{k}$
of the boundary circle and the vector $\hat{u}$ makes an obtuse
angle, so the geodesic curvature at any point on the boundary
circle is negative
$k_{g}=\vec{k}\cdot\hat{u}=-\frac{\sqrt{1-r^{2}}}{r}$, where $r$
is the radius of the boundary circle. A similar argument shows
that the sign of the geodesic curvature at the boundary of the
lower smaller part of the cut sphere in Fig.~\ref{cal_kg}(a) is
positive. Fig.~\ref{cal_kg}(b) lists all the possible shapes
around a symmetric circular pore of radius $r$ and the geodesic
curvature for each case. The first shape has the minimum geodesic
curvature, so it is preferred among other shapes. Therefore, the
Gaussian curvature term in the Helfrich free energy tends to pull
the membrane outside a pore to the plane where the pore loop lies.
This conclusion also holds for multi-pored membranes. From the
aspect of the Gauss-Bonnet theorem, the flattening effect of the
Gaussian curvature term is disclosed. It also sheds light on the
numerically generated flat surface in the vicinity of a pore on a
membrane when the Gaussian rigidity is tuned to be
negative.\cite{theory_of_PNAS:2005}

We use Surface Evolver to generate the ground state shape of a
pored membrane for exploring the flattening effect caused by the
Gaussian curvature term in the Helfrich free energy. The Surface
Evolver evolves a surface toward a local minimum energy shape by
calculating the force on each vertex from the gradient of the
total energy, which gives the direction of motion in the
membrane's configuration space.\cite{evolver} Therefore, the
method to generate a ground state shape by Surface Evolver is
distinct from that used in Ref.~\citenum{theory_of_PNAS:2005},
where the equilibrium shapes are produced from solving the shape
equation. The result is shown in Fig.~\ref{qiao} (b) for
$\kappa=2$ and $\kappa_{G}=-1.5$. A comparison of
Fig.~\ref{qiao}(a) and (b) shows that the Gaussian rigidity does
play a role in regulating the shape of a pored membrane. The mean
curvature term prefers to form a neck while the Gaussian curvature
term tends to flatten the membrane near the pore. A dark-field
micrograph of an experiment on a liposome with a pore whose size
(measured by the radius of the spherical body) is similar to that
in Fig.~\ref{qiao}(b) is shown in
Fig.~\ref{qiao}(c).\cite{talin:1998} The similarity of the shapes
in Fig.~\ref{qiao}(b, c) suggests that the experimental shape also
results from the competition of the mean curvature and the
Gaussian curvature terms.

The shape of the pore, as the result of the competition of the
mean curvature and the Gaussian curvature terms, encodes the
information about the ratio $\kappa_{G}/\kappa$, as has been
discussed in Refs.(~\citenum{Webb:2005, idema2011analytical,
semrau2006accurate}). Note that the absolute values of these
rigidities cannot be derived from the shape, because the shape is
determined only by their ratio. Here, we propose a scheme to
determine the quantitative relation between the shape of the pore
and the ratio $\kappa_{G}/\kappa$. Since the Gaussian curvature
term flattens the membrane near a pore, we approximate the shape
in Fig.~\ref{qiao}(b) as a combination of a circular truncated
cone (the section between the red line and the purple line) and a
spherical crown. The whole shape is characterized by three
parameters $r$, $A$, and $\theta$, where $r$ is the radius of the
pore, $A$ is the area of the membrane, and $\theta$ is defined in
Fig.~\ref{qiao}(b), which is referred to as \emph{the pore angle}.
The pore angle reflects the flatness of the membrane near the
pore. The total bending energy is
$E_{b}\left(r,A,\theta;\kappa_{G}/\kappa\right)
=\frac{1}{2}\kappa\int\left(2H\right){}^{2}dA+k_{G}2\pi\left(1+\cos\theta\right)$.
The mean curvature for sphere is $2H=2/R$ and for cone
$2H=\frac{\cos^{2}\delta}{z\sin\delta}$, where $2\delta$ is the
cone angle and $z$ is the vertical distance to the tip of the
cone. In $E_{b}\left(\theta,r,A;\kappa_{G}/\kappa\right)$, by
specifying $r$, $A$ (as measured from a given shape) and
$\kappa_{G}/\kappa$, we can find an optimal pore angle $\theta$
that minimizes the energy. We tune the ratio $\kappa_{G}/\kappa$
for fitting the optimal pore angle to the measured one. The ratio
$\kappa_{G}/\kappa$ is thus found from a given shape. This scheme
has its significance in application, considering that the ratio
$\kappa_{G}/\kappa$ is usually very difficult to measure in
experiment that only few results are
available.\cite{gaussian_rigidity:2012} On the other hand, the
scheme may be generalized to other systems, where the direct
measurement of the elastic moduli is difficult, like for living
materials.\cite{Fung_Biomechanics:1993,cell_bubble}

\begin{figure}[t]
\centering
\includegraphics[scale=1.0]{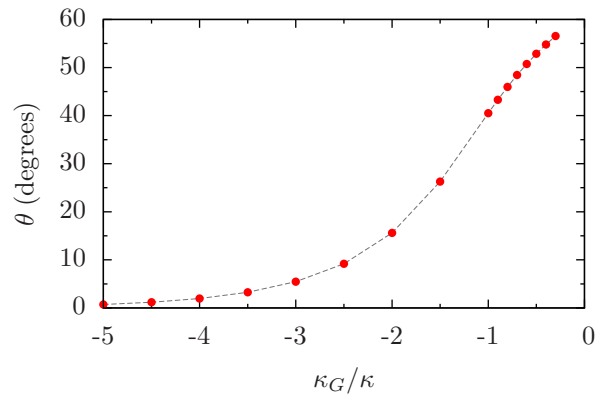}
\caption{The plot of optimal pore angle $\theta$ vs. the ratio of
$\kappa_{G}/\kappa$. The area is $62.8$ measured from the shape in
Fig.~\ref{qiao}(b) where the fixed radius of the pore is defined
to be unity. The membrane near the pore becomes more and more flat
($\theta$ decreases) with the increase of the absolute value of
$\kappa_{G}/\kappa$. For a real fluid membrane, $\kappa_G/\kappa
\in[-1,0]$, where more points are plotted. \label{theta_k} }
\end{figure}

We test the above method for finding the ratio $\kappa_{G}/\kappa$
of the shape in Fig.~\ref{qiao}(b). The radius of the pore is
defined as unity, so $R=2.3$ and the area is calculated as $62.8$.
By varying the ratio $\kappa_{G}/\kappa$, we get different optimal
pore angles, as shown in Fig.~\ref{theta_k}(a). It shows that the
membrane near the pore becomes more and more flat ($\theta$
decreases) with stronger flattening effect by the Gaussian
curvature term (the absolute value of $\kappa_{G}/\kappa$
increases). For fitting the optimal angle to the measured pore
angle $47^{\circ}$, the ratio is required to be
$\kappa_{G}/\kappa=-0.75$, which is exactly the one we use in
Surface Evolver to generate the shape in Fig.~\ref{qiao}(b). The
validity of the scheme for obtaining the ratio $\kappa_{G}/\kappa$
is thus substantiated.

Now we apply this scheme to the shape in Fig.~\ref{qiao}(c) for
identifying the ratio $\kappa_{G}/\kappa$ of the liposome used in
the experiment of Ref.~\citenum{talin:1998}. From the experimental
shape Fig.~\ref{qiao}(c), we measure $R=2.25$, pore angle
$\theta=55^{\circ}$ and calculate the area $A=66$. It is found
that the observed pore angle can be fitted by using
$\kappa_{G}/\kappa=-0.45$. Therefore, the value of the ratio
$\kappa_{G}/\kappa$ of the liposome in the experiment of
Ref.~\citenum{talin:1998} is estimated as $-0.45$, which is of the
same order as the experimentally-known values for typical
liposomes.\cite{gaussian_rigidity:2012}

\begin{figure}[t]
\centering{} \includegraphics[width=\columnwidth]{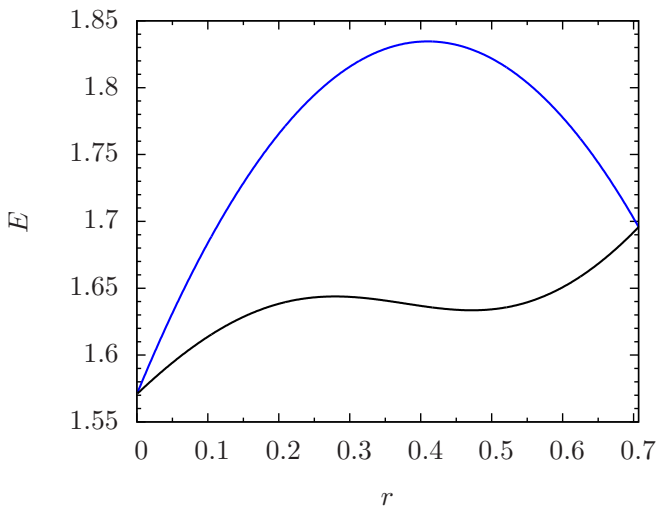}
\caption{The plot of energy versus the radius of the pore $r$ for
membranes with budding (black curve) and flat (blue curve) pores.
The two curves coincide at $r=\sqrt{2}/2 \approx 0.7$. It
corresponds to a hemisphere beyond which the ansatz shape of a
spherical cap plus catenoid does not apply. $\gamma=0.2049$.
$\kappa=1,\ \kappa_G=0,\ \textrm{A}=\pi$. The radius of the pore
$r$ is measured in the unit of the radius $r_0$ of the circular
disk whose area is fixed in the evolution. The meta-stable pore
has $r=0.43$, so the corresponding radius of the spherical cap is
$R=0.50$, $L=0.17$, and $\theta=68$ degrees. \label{line_tension}
}
\end{figure}

\subsection{Stability of a pore with line tension}

Finally, we briefly discuss the consequences of relaxing the
constraint of fixed pore size by introducing the line energy,
$\gamma \oint_{\partial} dl$ for the pore.~\cite{helfrich1974size}
We explore the stability of a budding pore by working in the
regime of $|\kappa_G|<<\kappa$ where a budding structure is
expected to form. The pored membrane is assumed to take the shape
of a spherical cap plus a catenoid, and the boundary of the pore
is approximated as falling on the waist of the catenoid. The
energy is thus obtained as
$E=\frac{\pi\kappa}{4}(1+\cos\theta)+\gamma 2\pi r+\kappa_G 2\pi$,
where $\theta$ is the pore angle and $r$ is the radius of the
pore. The area of the pore membrane is fixed: $\textrm{A}=\pi r
(2L+r\sinh(2L/r))+2\pi R^2(1+\cos\theta)$, where $L$ is the height
of the pore and $R$ is the radius of the spherical cap (see
Fig.~\ref{why_qiao}). $L$ and $R$ are related by
Eq.~(\ref{eq:LR}). For a given set of values for $\kappa,\
\kappa_G$ and $\textrm{A}$, the energy is a function of $r$ with
the free parameter $\gamma$. Fig.~\ref{line_tension} shows the
plot of the energy versus $r$ for pored membranes with budding
(black curve) and flat (blue curve) pores. The shape of a vesicle
with flat pore is approximated as a spherical
cap.~\cite{helfrich1974size} Fig.~\ref{line_tension} shows that
for a specified value for the line tension the pore vanishes in
both cases in the ground state. We notice that a budding pore has
a meta-stable state at about $r=0.43$. However, this meta-stable
state may be hard to see in an experiment, because the depth of
the energy barrier ($\sim 0.01\kappa$) is very shallow and the
range of values of the line tension where a meta-stable pore
exists is very narrow: $0.19\lesssim r_0\gamma/\kappa \lesssim
0.22$, where $r_0$ is the radius of the circular disk as defined
in the caption of Fig.~\ref{line_tension}. We perform a series of
simulations using Surface Evolver by adding the line energy to the
pore. We were not able to observe a stable pore, \textit{i.e.},
the pore either shrinks and closes up (for large values of line
tension) or it fully opens and the membrane takes a form of a flat
disk (for small line tension). While our numerical results cannot
exclude the possibility of the existence of a stable pore within a
certain parameter region, they suggest that even if such region
exists, it is very narrow. Therefore, stabilizing agents like
talin proteins in the experiment of Ref.~\citenum{talin:1998} are
essential for a stable pore on fluid membranes.

\subsection{Two-component membrane}

\begin{figure}[t]
\centering{}  \includegraphics[width=1.4in]{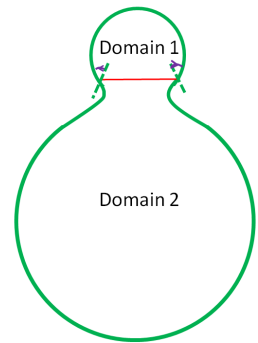} \caption{The
schematic plot of budding on a two-component membrane. The red
line represents the boundary of the two domains.
$|\kappa_{G}^{1}|>|\kappa_{G}^{2}|$.\label{two_comp} }
\end{figure}

So far we have studied the effects of the mean curvature and the
Gaussian curvature terms in the Helfrich free energy on the shape
of pored membranes. It is interesting to extend the flattening
effect due to the Gaussian curvature to two-component membranes
where the components' Gaussian rigidities are different. A pored
membrane may be regarded as a limiting case of a two-component
membrane, where one phase has vanishing bending and Gaussian
rigidities. The effect of the inhomogeneity of the Gaussian
rigidity in multicomponent membranes has been extensively
discussed.\cite{Webb:2005, allain2004fission, idema2011analytical,
julicher1993domain, Hu2011} Monte Carlo simulations show that a
difference in the Gaussian rigidity of a two-component membrane
can develop and stabilize multi-domain
morphologies.\cite{julicher1993domain,Hu2011,allain2004fission} An
explicit analytical expression for the shapes of axisymmetric
closed membranes with multiple domains is derived in
Ref.~\citenum{idema2011analytical}. However, the influence of the
inhomogeneity of the Gaussian curvature on the local shape near
the phase boundary was not explicitly discussed. In this
subsection, we study how the same Gaussian-curvature effect that
leads to the flattening near a pore can result in the onset of
budding in a multicomponent membrane, if the Gaussian rigidities
of the components are different. For simplicity, consider a
two-component spherical membrane with Gaussian rigidities
$\kappa_{G}^{(1)}$ and $\kappa_{G}^{(2)}$ for domain 1 and domain
2 of the sphere, respectively (see Fig.~\ref{two_comp}). Suppose
$\Delta\kappa_{G}=\kappa_{G}^{(2)}-\kappa_{G}^{(1)}>0$ without
loss of generality. The integral of the Gaussian curvature over
the whole surface is
$\kappa_{G}^{(1)}\int_{1}K_{G}dA+\kappa_{G}^{(2)}\int_{2}K_{G}dA=
2\pi\left(\kappa_{G}^{(1)}+\kappa_{G}^{(2)}\right)-
\Delta\kappa_{G}\oint_{2}k_{g}dl =
2\pi\left(\kappa_{G}^{(1)}+\kappa_{G}^{(2)}\right)+\Delta\kappa_{G}\oint_{1}k_{g}dl$,
where the subscript numbers in the line integrals represent the
boundary of the respective domains. The second and third
expressions indicate that the geodesic curvature on the boundary
of domain 2 (with larger Gaussian rigidity) prefers to increase
and that on the boundary of domain 1 (with smaller Gaussian
rigidity) prefers to decrease for lowering the Helfrich free
energy. The effect is similar to imposing a ``torque'' rotating
outward the original shape near the boundary loop (the dashes
lines in Fig.~\ref{two_comp}).

\begin{figure}[t]
\centering{} \includegraphics[scale=1.0]{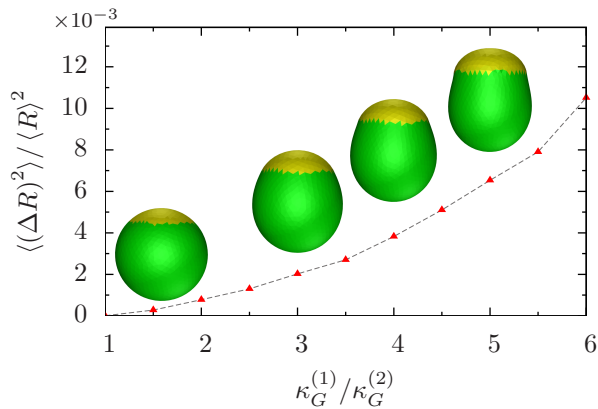} \caption{The
plot of the asphericity of a two-component membrane vs. the ratio
of the two Gaussian rigidities in the two-component membrane with
the 15\% of the purple (domain 1) component. The budding becomes
more obvious with the increase of the inhomogeneity of the
membrane in Gaussian rigidity. \label{kG12} }
\end{figure}

In order to confirm the proposed budding scenario, we performed a
series of simulated annealing Monte Carlo simulations for a
triangulated two-component membrane. Components were assigned to
the vertices of the discrete mesh and liquid character of the
membrane is ensured by using a dynamical triangulation; {\em
i.e.}, we employed a Monte Carlo move in which an edge shared by
two triangles was flipped to connect two vertices that were
previously not connected.\cite{kazakov1985critical,SMoMS} The
discrete version of the mean curvature term in the Helfriech free
energy was calculated following a prescription introduced by
Gompper and Kroll,\cite{gompper1996random} while the Gaussian
curvature term was treated according to Meyer, \emph{et
al}.\cite{meyer2002discrete} For a membrane with about
$2\times10^{3}$ vertices typically $10^{5}$ Monte Carlo sweeps
with a linear cooling protocol was sufficient to obtain low energy
structures, with a sweep defined as an attempted move of each
vertex followed by an attempted flip of each edge.

The result is shown in Fig.~\ref{kG12}. In the simulation,
$\kappa=2$, $\kappa_{G}^{(2)}=-0.5$ and
$\kappa_{G}^{(1)}/\kappa_{G}^{(2)}$ increases from unity to 6. The
deviation from a spherical shape is characterized by the
asphericity $\frac{<(\Delta
R)^{2}>}{<R>^{2}}=\frac{1}{N}\sum_{i=1}^{N}\frac{(R_{i}-<R>)^{2}}{<R>^{2}}$,
where $R_{i}$ is the radial distance of vertex $i$ and
$<R>=\frac{1}{N}\sum_{i=1}^{N}R_{i}$ is the mean
radius.\cite{Lidmar03} With the increasing inhomogeneity in the
Gaussian rigidity, the ``torque'' imposed on the phase boundary
becomes stronger and the budding of the smaller component becomes
more obvious as shown in Fig.~\ref{kG12}. This budding mechanism
arising from an inhomogeneity in the Gaussian rigidity is distinct
from the usual mechanism due to line tension. It sheds light on
understanding shapes of multicomponent membranes and provides a
novel method to control the shape of membranes.

\section{Conclusions}

Our study of shapes of pored membranes of fixed area and pore size
within the framework of the Helfrich theory shows that the
presence of pores can be an important ingredient for generating
various shapes of membranes. Several structures brought by pores
have been disclosed, including the budding pores purely due to the
mean curvature term and the flattening effect due to the Gaussian
curvature term. The latter effect may be used to fabricate
pore-controlled buckled membranes. Furthermore, we have proposed a
method to extract the value of the Gaussian rigidity of a membrane
simply from its shape. This scheme may be generalized to systems
where the elastic moduli are difficult to measure, like in living
materials. In addition, by relaxing the constraint of a fixed pore
size and adding the line tension, we briefly discuss the stability
of a pore and find that a budding pore may be meta-stable with
very shallow energy barrier within a narrow range of line tension
values. Finally, we extend the flattening effect due to the
Gaussian curvature as found in studying pored membranes to
two-component membranes. Theoretical analysis shows that
sufficiently high contrast between the components' Gaussian
rigidities can lead to budding of a two-component membrane, which
is substantiated by MC simulations.

Numerical simulations were in part performed using the
Northwestern University High Performance Computing Cluster Quest.
ZY, MO and CT thank the financial support of the Air Force Office
of Scientific Research (AFOSR) under Award No. FA9550-10-1-0167.
RS and MO thank the financial support of the US Department of
Energy Award DEFG02-08ER46539.

\providecommand*{\mcitethebibliography}{\thebibliography}
\csname @ifundefined\endcsname{endmcitethebibliography}
{\let\endmcitethebibliography\endthebibliography}{}

\end{document}